\pdfoutput=1

\documentclass{article}[12pt]
\usepackage{fullpage}
\usepackage[utf8]{inputenc}
\usepackage{times}
\usepackage{inconsolata} 
\usepackage[svgnames,x11names]{xcolor}
\usepackage{hyperref}
\usepackage{listings}
\usepackage{todonotes}
\usepackage{amsmath}
\usepackage{booktabs}
\usepackage{subcaption}
\usepackage{amssymb}
\newtheorem{definition}{Definition}

\makeatletter
\define@key{todonotes}{Mihai}[]{%
    \setkeys{todonotes}{author=\textbf{Mihai:},color=green}}%
\define@key{todonotes}{Somesh}[]{%
    \setkeys{todonotes}{author=\textbf{Somesh:}}}%
\makeatother

\title{Formal Analysis of the API Proxy Problem}
\author{Somesh Jha, Mihai Christodorescu, Anh Pham}
\date{} 

\begin{document}

\maketitle

\begin{abstract}
    Implementing a security mechanism on top of APIs requires clear understanding of the semantics of each API, to ensure that security entitlements are enforced consistently and completely across all APIs that could perform the same function for an attacker. Unfortunately, APIs are not designed to be "semantically orthogonal" and they often overlap, for example by offering different performance points for the same functionality. This leaves it to the security mechanism to discover and account for \textit{API proxies}, i.e., groups of APIs which together approximate the functionality of some other API. Lacking a complete view of the structure of the API-proxy relationship, current security mechanisms address it in an ad-hoc and reactive manner, by updating the implementation when new API proxies are uncovered and abused by attackers.
    
    We analyze the problem of discovering API-proxy relationships and show that its complexity makes it NP-complete, which makes computing exact information about API proxies prohibitively expensive for modern API surfaces that consist of tens of thousands of APIs. We then propose a simple heuristic algorithm to approximate the same API-proxy information and argue that this overapproximation can be safely used for security purposes, with only the downside of some utility loss. We conclude with a number of open problems of both theoretical and practical interest and with potential directions towards new solutions for the API-proxy problem.
\end{abstract}


\section{Introduction}
\label{sec:intro}


Enforcement of a security or privacy policy is often achieved through restrictions on the application programming interface (API) a system presents to applications, typically by having calls from an application to platform APIs\footnote{For simplicity we will use the term ``API'' to denote a function that a system exposes to apps and users.} be protected by a check for the corresponding permission, capability, or related security entitlement. This allows one to enforce a security policy by blocking an API for a sandboxed application or by making an API available only to applications run by privileged users. Such enforcement requires complete mediation, to guarantee that every access to every object is protected by a security check~\cite{Ref-saltzer-1975}. Unfortunately the design of modern platform APIs,
such as Android, do not lend itself to complete mediation, as oftentimes multiple APIs to accessing the same information or functionality are present in order to enable different performance points and use cases or to maintain backward compatibility. In turn this places the burden on the security mechanism to determine if the protection applied to an API must be extended to other APIs. Complete mediation for API-based security requires understanding and handling such API substitutes or proxies.

Determining if an API is substitutable by other APIs, or conversely if a group of APIs can serve as proxies for a given API, is challenging. Consider the case of Apple's App Tracking Transparency (ATT) program for iOS, which aims to reduce user tracking by apps by protecting user-id APIs behind consent prompts. After this program was deployed, subsequent studies showed that some apps evaded the ATT protection by using APIs that access attributes serving as substitutes for user-id, including locale, mobile carrier, timezone, system disk space, and system version. Thus securing one API but ignoring all of its proxies has resulted in a weakly enforced ATT program. As we described in \autoref{sec:examples}, there exist two types of API proxies, distinct APIs that access the same platform attribute (e.g., locale and country APIs both retrieve the geographical-region setting) and distinct APIs that access different attributes which are correlated to the same external, real-world attribute (e.g., cellular-network cell identifier and WiFi access point both reflect real-world location). Thus one challenge of computing the API-proxy relationship is due to the lack of observability into the confounding attributes that cause APIs to be substitutable.

Efforts to handle API proxies to allow the implementation of trustworthy security mechanisms have been fragmented at best and have taken one of two paths. One approach is to design the security mechanism to take into account these API proxies. The Android permission to access location is one such example, as it covers not only access to GPS-provided location, but also to information about WiFi and Bluetooth, both of which can act as proxies for GPS location. (Note that this permission is not comprehensive, as it does not cover cellular network or TCP/IP information.) Another approach is to identify the confounding attributes that are common to multiple APIs and then move the enforcement mechanism to that layer of the system, instead of API layer that exhibits the API proxies. In SELinux policies, security labels (which define which accesses will be allowed and which will be denied) are defined with respect to fully qualified file paths, but enforce at the level of file objects on disk, irrespective of the actual file path taken to reach that file object. As a result, although most APIs for file access are defined in terms of file paths, the security enforcement is placed at the level of file objects, thus avoiding API proxying due to multiple file paths pointing to the same file object (e.g., via POSIX-style filesystem links). In both of these approaches, the solutions are ad-hoc and custom to selected platform APIs and attributes and do not generalize to the full API set. Thus a second challenge is that of discovering and characterizing the API-proxy relationship automatically and at scale, over the whole API set (as a data point, Android has some 35,000+ APIs).

Our approach to understanding the API proxy problem is two-fold. First we analyze the computational complexity of the problem, to determine a bound on the cost of identifying API proxies in terms of the size of attributes and APIs that exist on the platform. Second we propose a greedy algorithm for identifying API proxies, which is amenable to practical implementation but may fall short of the exact solution (in other words, it may output an API proxy set over-approximating the true set for a given platform attribute).


In \autoref{sec:formalization} we formalize the {\it API Proxy Problem (ApiPP)} and analyze its complexity.
The problem instance consists of a set of finite attributes and a distribution over the universe of attribute
values. Moreover, we are a given a specific attribute (e.g. location) and a set of functions, where each function takes a subset of the attributes and produces a value (in the context of Android, these
functions can be thought of as APIs). The decision version of ApiPP asks whether there exists 
a subset of functions or APIs that reduce the uncertainty of the specified attribute (i.e. given the values of the functions the specified attribute can be predicted with uncertainty below a certain threshold). In other words,
these functions can be used as a proxy for that attribute. We also prove that
ApiPP problem is NP-complete (the reduction is from the classic vertex cover problem). We note that similar "proxy problems" appear in other contexts, such as fairness in machine learning. In fairness in machine learning, the proxy problem corresponds to whether there is a set of features that can predict well the sensitive attributes (e.g. gender and race). Exploring connections between proxy problems in different contexts is an interesting area of future
research.


The rest of the paper is organized as follows. In \autoref{sec:examples} we provide detailed examples of API proxies from two domains and their security implications. The complexity of the API-proxy decision problem is formalized and shown to be NP-complete in \autoref{sec:formalization}, with a heuristic solution presented in \autoref{sec:heuristics}. We then discuss the implications of this result and describe a series of open question in \autoref{sec:discuss}.


\section{Examples}
\label{sec:examples}

We illustrate the problem of API proxies with examples from the Android and iOS operating systems (OSes). Both of these OSes manage a large number of data items (ranging from user data, to system settings, and to hardware-sensor outputs) and offer a correspondingly large collection of APIs that apps can use to retrieve such data items and compute over them. Our examples highlight a small part of the problem space and for each example we show the data items, system settings, or sensor outputs in consideration, the APIs that operate over such data, and the API proxies given as a group of APIs together with evidence that this proxy relationship exists. For brevity we will refer to all data items, system settings, and sensor outputs as ``platform attributes'' or simply ``attributes.''


\subsection{API Proxies for Display Size in Android}

Consider the display size of an Android device. Given that the display size correlates with the price of the device and thus with the socioeconomic status of the user, the display size can often be a sensitive piece of information. A user may wish to control access to such sensitive data by setting a policy that blocks the \lstinline!Display.getMetrics! API unless explicitly consented by the user. There are a number of adjacent APIs that unfortunately proxy the same display-size information as \lstinline!Display.getMetrics!.

\begin{center}
\begin{tabular}{ll}
\toprule
    \textit{Android API} & \textit{Description} \\
\midrule
    \lstinline!Display.getMetrics!     & retrieve the display height, width, density, and font-scaling factor \\
    \lstinline!Activity.getWindow!     & get the current app window \\
    \lstinline!Window.setFlags!        & set layout options for a window \\
    \lstinline!Window.setContentView!  & place a view in a window \\
    \lstinline!View.getWidth!          & get the width of a view \\
    \lstinline!View.getHeight!         & get the height of a view \\
\bottomrule
\end{tabular}
\end{center}

In particular an app can avoid any direct calls to \lstinline!Display.getMetrics! and instead use the dimensions of a view in a maximized window to infer the actual dimensions of the display. Thus we land on the following API proxies for the \textit{display-size} attribute, as shown in \autoref{t:android_display_examples}.

\begin{table}
    \centering
    \subcaptionbox{API proxy group \# 1}[0.4\textwidth]{
        \begin{tabular}{l}
            \toprule
            \lstinline!Display.getMetrics! \\
            \bottomrule
        \end{tabular}
    }
    \subcaptionbox{API proxy group \# 2}[0.4\textwidth]{
        \begin{tabular}{l}
            \toprule
            \lstinline!Activity.getWindow!    \\
            \lstinline!Window.setFlags!       \\
            \lstinline!Window.setContentView! \\
            \lstinline!View.getWidth!         \\
            \lstinline!View.getHeight!        \\
            \bottomrule
        \end{tabular}
    }
    \caption{APIs that can be proxies for the \textit{display-size} attribute.}
    \label{t:android_display_examples}
\end{table}

We note that these groups of APIs are equivalent for the purpose of obtaining the display size, though the API proxy group \# 2 does not retrieve the additional information that \lstinline!Display.getMetrics! provides (display density and display font-scaling factor).


\subsection{API Proxies for Location in Android}

Consider the following attributes available on the Android OS: \textit{location} (representing geographical location as a $\langle\textit{latitude}, \textit{longitude}\rangle$ tuple), \textit{WiFi-list} (the list of nearby WiFi access points), \textit{cell-list} (the list of nearby cellular towers), \textit{BT-list} (the list of nearby Bluetooth devices), and \textit{IP-list} (the list of TCP/IP addresses assigned to the network interfaces). The Android platform provides APIs to retrieve these attributes and refresh them on demand, and these attributes are used both by Android built-in services and by Android apps. The relevant APIs are listed below: 
\begin{center}
\begin{tabular}{ll}
\toprule
    \textit{Android API} & \textit{Description} \\
\midrule
    \lstinline!LocationManager.getCurrentLocation!     & retrieve the current location from GPS \\
    \lstinline!WifiManager.getScanResults!             & get the list of WiFi access points in proximity \\
    \lstinline!TelephonyManager.getAllCellInfo!        & get the list of cellular network towers in proximity \\
    \lstinline!BluetoothAdapter.startDiscovery!        & discover Bluetooth devices in proximity \\
    \lstinline!NetworkInterface.getInterfaceAddresses! & get the list of addresses for a network interface \\
\bottomrule
\end{tabular}
\end{center}

Many WiFi access points, GSM cells, TCP/IP addresses, and Bluetooth beacons are associated with a geographical location and databases mapping the radio devices and network addresses to geographical locations exist. Then one can approximate the geographical location by using the assigned TCP/IP addresses together with identifiers of neighboring radio devices (for WiFi, cell, and/or Bluetooth) and their associated signal strength. Indeed functionality in the Android OS (partially) builds on this observation to offer location information to apps while also optimizing the battery usage (since each of these APIs has different power-consumption patterns)~\cite{android-fused-location}:
\begin{quote} \it
    The fused location provider is a location API in Google Play services that intelligently combines different signals to provide the location information that your app needs.
    
    The fused location provider manages the underlying location technologies, such as GPS and Wi-Fi, and provides a simple API that you can use to specify the required quality of service. For example, you can request the most accurate data available, or the best accuracy possible with no additional power consumption.
\end{quote}
\autoref{t:android_proxy_examples} lists the groups of APIs that act as proxies for the \textit{location} attribute.

We note that these groups of APIs are not equivalent, as location information from the \lstinline!LocationManager! is the most precise while signal strengths of nearby radio devices give only an approximation of the same location information. Another observation is that this approximation varies by location and thus is best described by a probability distribution over the domain $\textit{location} \times \textit{WiFi-list} \times \textit{cell-list} \times \textit{BT-list} \times \textit{IP-list}$. This distribution arises from the fact that not all geographical locations have WiFi service, cellular service, and Bluetooth beacons. Indeed in many cases only one or two of these services exist and thus the accuracy of the approximation varies accordingly.

\begin{table}
    \centering
    \subcaptionbox{API proxy group \# 1}[0.4\textwidth]{
        \begin{tabular}{l}
            \toprule
            \lstinline!LocationManager.getCurrentLocation! \\
            \bottomrule
        \end{tabular}
    }
    \subcaptionbox{API proxy group \# 2}[0.4\textwidth]{
        \begin{tabular}{l}
            \toprule
            \lstinline!WifiManager.getScanResults!             \\
            \lstinline!TelephonyManager.getAllCellInfo!        \\
            \lstinline!BluetoothAdapter.startDiscovery!        \\
            \lstinline!NetworkInterface.getInterfaceAddresses! \\
            \bottomrule
        \end{tabular}
    }
    \caption{APIs that can be proxies for the \textit{location} attribute.}
    \label{t:android_proxy_examples}
\end{table}

The security implication of these API proxies is to require a permission that covers all APIs within these proxy groups. Android does indeed take this route by defining permissions for two levels of precision, \lstinline!ACCESS_COARSE_LOCATION! and \lstinline!ACCESS_FINE_LOCATION!, and mandating that apps obtain one of these permissions before accessing \textit{some} of the APIs in the proxy groups.


\subsection{API Proxies for User Identity in iOS}
\label{sec:ios-examples}

Consider the following attribute available on the iOS OS: \textit{user-id} (representing the identity of the device owner). The iOS platform maintains this ID as a way to connect the owner of the device to the corresponding Apple cloud account. At the same time, having this \textit{user-id} available directly to apps leads to a number of privacy concerns, ranging from directly leaking personally identifiable information (PII) to third-party app developers, to enabling tracking of a user's activity across apps. The iOS platform tries to protect against the problem of PII leakage by shielding the \textit{user-id} behind an Identifier for Advertisers (IDFA), available to apps via the \lstinline!ASIdentifierManager.advertisingIdentifier! API. The problem of cross-app tracking was addressed by Apple's App Tracking Transparency program (ATT), which restricted the app access to IDFA by requiring user consent.

Kollnig et al.~\cite{10.1145/3531146.3533116} studied the impact of the ATT program on apps that previously retrieved the IDFA and discovered that some apps actively sought to use alternative API alternative to obtain equivalent information. Let us consider the following set of iOS APIs corresponding to platform attributes \textit{free-space}, \textit{total-space}, \textit{mobile-carrier}, \textit{country}, \textit{locale}, \textit{CPU-arch}, \textit{timezone}, \textit{phone-model}, \textit{phone-name}, and \textit{os-version}. In most cases the APIs are simple wrappers that return the value of the corresponding attribute, without any additional processing, as shown in \autoref{t:ios-apis}. The study indicates that, based on the evolution of the apps analyzed, at least two subsets of APIs are each proxies for user identity, just as \lstinline!ASIdentifierManager.advertisingIdentifier! is. We summarize this information in \autoref{t:ios_proxy_examples}.

\begin{table}
    \centering
    \begin{tabular}{ll}
    \toprule
        \textit{iOS API} & \textit{Description} \\
    \midrule
        \lstinline!FileManager.attributesOfFileSystem!      & free and total disk space \\
        \lstinline!CTCarrier.carrierName!                   & name of mobile service provider \\
        \lstinline!NSLocale.currentLocale.countryCode!      & country code \\
        \lstinline!NSLocale.currentLocale.localeIdentifier! & locale identifier \\
        \lstinline!NXGetLocalArchInfo.description!          & CPU architecture \\
        \lstinline!TimeZone.current!                        & timezone \\
        \lstinline!UIDevice.identifierForVendor!            & ID for vendors \\
        \lstinline!UIDevice.localizedModel!                 & phone model \\
        \lstinline!UIDevice.name!                           & user-given phone name \\
        \lstinline!UIDevice.systemVersion!                  & OS version \\
    \bottomrule
    \end{tabular}
    \caption{iOS APIs from the study on ATT evasion by Kollnig et al.~\cite{10.1145/3531146.3533116}.}
    \label{t:ios-apis}
\end{table}

\begin{table}
    \centering
    \subcaptionbox{API proxy group \# 1}{
        \begin{tabular}{l}
            \toprule
            \lstinline!ASIdentifierManager! \\
            \qquad \lstinline!.advertisingIdentifier! \\
            \bottomrule
        \end{tabular}
    }
    \subcaptionbox{API proxy group \# 2}{
        \begin{tabular}{l}
            \toprule
            \lstinline!CTCarrier.carrierName! \\
            \lstinline!NSLocale.currentLocale! \\
            \qquad \lstinline!.localeIdentifier! \\
            \lstinline!NXGetLocalArchInfo.description! \\
            \lstinline!TimeZone.current! \\
            \lstinline!UIDevice.localizedModel! \\
            \lstinline!UIDevice.name! \\
            \lstinline!UIDevice.systemVersion! \\
            \bottomrule
        \end{tabular}
    }
    \subcaptionbox{API proxy group \# 3}{
        \begin{tabular}{l}
            \toprule
            \lstinline!FileManager.attributesOfFileSystem! \\
            \lstinline!NSLocale.currentLocale.countryCode! \\
            \lstinline!TimeZone.current! \\
            \lstinline!UIDevice.identifierForVendor! \\
            \lstinline!UIDevice.localizedModel! \\
            \lstinline!UIDevice.name! \\
            \lstinline!UIDevice.systemVersion! \\
            \bottomrule
        \end{tabular}
    }
    \caption{iOS APIs that can be proxies for \textit{user-id}~\cite{10.1145/3531146.3533116}.}
    \label{t:ios_proxy_examples}
\end{table}

The goal of the apps using the API proxies is not to reconstruct the exact value of the \textit{user-id} attribute, but rather to obtain a set of attribute values that correlate to \textit{user-id}. In other words, an app using one of the API proxies would wish to derive an alternative ID that is with high probability correlated to a single \textit{user-id}. Similar to the Android example, there exists a non-trivial distribution over the joint attribute space $\textit{user-id} \times \textit{free-space} \times \textit{total-space} \times \textit{mobile-carrier} \times \dots$, reflecting the real-world demographics of the iOS user base (there are many users with the same locale, many users with the same mobile carrier, same disk space, etc.). Using multiple APIs as a proxy is a way to reduce the uncertainty of the predicted \textit{user-id} by combining multiple random variables.

The security implication of these API proxies is that iOS design now has to decide whether all of the above APIs should be restricted behind a user-consent prompt. Since some of these APIs can be used for other, benign purposes, this introduces a hard trade-off between security and usability (as the user would be required to evaluate app requests to use such APIs before providing consent).




\section{Formalization}
\label{sec:formalization}

Let $\mathcal{A} = \{a_1,\cdots.a_m \}$ be a finite set of attributes (e.g., location, device-id). Each attribute $a_i$
($1 \leq i \leq m$) is associated with a finite domain $D_i$. Let $\mathcal{D}(\mathcal{A})$ be a a distribution over
the joint domain of attributes (i.e., distribution over $D_1 \times \cdots \times D_m$). Let $Samp(\mathcal{D}(\mathcal{A}))$ be a sampling oracle over $\mathcal{D}(\mathcal{A})$.
Let $F=\{ f_1,\cdots,f_n \}$ be set of functions (each function can be thought of as 
an API in the context of Android). Each function $f_i$ takes values for a set of attributes
$A_i \subseteq A$ and produces values in the domain $Z_i$.

\begin{definition}[API proxy decision problem (ApiPP)]
\rm
Given an attribute $a \in \mathcal{A}$, a set of functions $F=\{ f_1,\cdots,f_n \}$, a distribution of attribute values $Samp(\mathcal{D}(\mathcal{A}))$, an uncertainty estimator $U$, and a bound $k \leq n$, find a subset of functions $F' \subseteq F$ of size at most $k$, $| F' | \leq k$, such that $U(a,F') \leq \alpha$.
\end{definition}

Informally, the function $U(a,F')$ says what the ``uncertainty'' of the attribute $a$ is given,
values in the function $F'$ (e.g., if $F'$ contains a function that simply gives the value of the attribute $a$,
then uncertainty is $0$). Let $n$ be the length of the representation of the problem instance (e.g., attributes $a_i$, domains $D_i$, functions $f_i$). We require that $U(\cdot,\cdot)$ be computed in time polynomial in $n$.
Each call to the sampling oracle to obtain a new sample from the distribution has cost $O(1)$. We call the above problem the {\it local API proxy decision problem (ApiPP)}.

We show that \textit{ApiPP} is NP-complete. Our reduction will be from vertex cover problem. Given an undirected graph $G=(V,E)$, where $E \subseteq V \times V$, $V' \subseteq V$ is called a vertex cover if every 
edge $E$ has endpoint in $V'$. The {\it k-VP} problem asks if $G$ has a vertex cover of
size $\leq k$. 
Given a graph $G=(V,E)$ let $n$ and $m$ be the size of $V$ and $E$, respectively.
The set of attributes $A=\{ a_1,\cdots,a_n,a_{n+1} \}$, where each attribute is 
binary (i.e. takes value $0$ and $1$). Attribute $a_i$ ($1 \leq i \leq n$) corresponds to whether the $i$-th vertex is present or not. Attribute $a_{n+1}$ is
a special attribute, which we will describe next. Distribution $\mathcal{D}(\mathcal{A})$ is described as follows: 
value of first $n$ attributes are picked uniformly randomly, and if these values corresponding to a set cover for $G$, then $a_{n+1}$ is $1$ with
probability $1$, and thus $0$ with probability $0$; otherwise $a_{m+1}$ is uniformly random (i.e. value $0$ or $1$ with probability $0.5$). The structure of the graph
$G$ is implicitly encoded in the distribution. 
Set of functions $F=\{ f_1,\cdots , f_n \}$ is defined as follows: $f_i$ takes
attribute $a_i$ and simply returns its value (essentially it is encoding of
the $i$-th vertex). The special attribute $a$ is
$a_{n+1}$. For $F' \subseteq F$ we define $U(a_{n+1},F')$ as entropy of 
$a_{n+1}$ conditioned on that the  attributes corresponding to $F'$ are set to $1$ and others are set to $0$ (i.e. if $F'$ corresponds to a cover the entropy is $0$ otherwise  it is $1$). Note that $U (a_{n+1},F') \leq 0.5$ iff $F'$ corresponds
to a vertex cover, and thus reduction is complete.
Since \textit{ApiPP} is in NP and is NP-hard, it is NP-Complete. $\square$

\vspace{\baselineskip}
We can also define the related optimization problem \textit{min-ApiPP}. 

\begin{definition}[API proxy optimization problem (min-ApiPP)]
\rm
Given an attribute $a \in \mathcal{A}$, a set of functions $F=\{ f_1,\cdots,f_n \}$, a distribution of attribute values $Samp(\mathcal{D}(\mathcal{A}))$, and an uncertainty estimator $U$, find a subset of functions $F' \subseteq F$ of minimum size such that $U(a,F') \leq \alpha$.
\end{definition}


\section{A Heuristic Solution to \textit{min-ApiPP}}
\label{sec:heuristics}


Let us taken an instance of \textit{min-ApiPP},
$$\langle a, F, Samp(\mathcal{D}(\mathcal{A})), U, \mathcal{A} \rangle,$$
with functions $F =\{ f_1,\cdots,f_n \}$ and attribute $a \in \mathcal{A}$.


We construct a greedy algorithm to approximate \textit{min-ApiPP} by taking
advantage of the fact that the uncertainty estimator $U$ is monotonically decreasing
in its second argument, because the uncertainty value $U(a,F')$ does not increase as $F'$
grows to include more functions.
Let $X(f_i)$ be the random variable corresponding
to function $f_i$ induced by the distribution $\mathcal{D}(\mathcal{A})$. Given
a subset of functions $F' \subseteq F$, we can define $U$ as:
$$U(a,F') = I(a , \{ X_f \}_{f \in F' }),$$
where $I(\cdot,\cdot)$ is the mutual-information function. We seek
$F'$ such that $I(a , \{ X_f \}_{f \in F' })$ is the
less than or equal to $\alpha$.
The greedy algorithm starts by picking a function $f_1 \in F$ that
maximizes $I(a,X_{f_1})$ (i.e., initially $F'=\{ f_1 \}$). Given $F'$, we add a function 
$f \in F \setminus F'$ that maximizes $I(a , \{ X_f \} \cup \{ X_g \}_{g \in F'})$ and we 
keep adding functions to $F'$ until $I(a , \{ X_f \}_{f \in F' })$ is
less than or equal to $\alpha$. The mutual information $I$ can
be estimated empirically by sampling by $Samp(\mathcal{D}(\mathcal{A}))$. Note that this algorithm has time complexity $O(n^2)$.


\section{Discussion}
\label{sec:discuss}

With the above formalism in place, we consider the implications for the practical design of security mechanisms on top of APIs in present-day systems. We also provide a list of open problems of both theoretical and practical interest.


\subsection{Implications}

The NP-completeness of \textit{ApiPP} makes the precise enforcement of a security policy, even for a single API, impractically expensive in the general case. 
There are two reasons for this.

First, treating APIs as blackboxes and trying to brute-force the discovery of API proxies is not realistic for a large API surface. The combinatorial explosion from having to consider all subsets when searching for API proxies means that API surfaces of present-day systems cannot be analyzed in full.

Second, using a heuristic to discover API proxies within some approximation bound has to be done safely so that the resulting information can be used for complete security enforcement. An underapproximate API proxy set, which lacks some APIs, is not safe for enforcement as it leaves opportunities for the adversary to evade the security enforcement. Our greedy algorithm in \autoref{sec:heuristics} computes an overapproximation of the API proxy set for a target API, and thus is safe for security purposes, but can induce a loss of utility if the API proxy set is overly large and the security mechanism restricts too many APIs. We discuss in the next subsection the open problem of balancing safety and utility in a heuristic algorithm.

Another key observation is that \textit{ApiPP} is parametrized by a distribution $\mathcal{D}(\mathcal{A})$ over the joint domain of attributes. This distribution is not a theoretical artifact, but rather captures the fact that attributes often reflect real-world properties and thus are not uniformly distributed. This means that API proxies can exist even for seemingly unrelated APIs simply because the underlying real-world properties happen to be correlated. This also means that, as API proxies are defined only within the distribution of attribute values, when the distribution changes, the API proxies have to be rediscovered.


\subsection{Open Problems}
\label{sec:open}

\paragraph{ApiPP-aware enforcement.}
Even with a satisfactory solution to the ApiPP, approximate or exact, it is unclear how to use the API-proxy information for an enforcement mechanism such as API blocking, sandboxing, or misuse detection. Blocking an API and all of its proxy sets may provide the best security guarantee for enforcement, but at the same time it may be overly strict.

It is possible that blocking a small number of APIs would prevent an adversary from making use of a proxy set (because they can no longer achieve a low-enough uncertainty level less than $\alpha$), without blocking all of the APIs in the proxy set. Determining which APIs to block to achieve effective security is specific to the security policy of interest: for a data-leakage policy, even partial leakage may be undesirable ($\alpha$ is high) and thus all APIs in the proxy set would need to be blocked; for a system-integrity policy, an incomplete execution of the APIs in the proxy set will fail to achieve the integrity breach ($\alpha$ is low) and thus blocking one or a few APIs will suffice.

\paragraph{Global ApiPP.}
Recall that our formalization of the \textit{ApiPP} attempt to find a subset of functions $F' \subseteq F$
that reduces the uncertainity of one attribute $a \in A$. We can define the \textit{global version of the
ApiPP} where we attempt to find proxies for a set of attributed $A' \subseteq A$. Note that the 
global \textit{ApiPP} problem is much harder than the problem for one attribute, but perhaps is more realistic
(e.g. find the proxies of all sensitive attributes, such as location and device-id). There is also 
opportunity that finding proxies for one attribute might help in finding proxies of another attribute. Exploring
the global proxy problem and finding efficient algorithms for it is an interesting open problem.

\paragraph{Improved approximation for local, min, and global ApiPP.}
Recall that we proved that \textit{ApiPP} is NP-hard using a reduction from the classic vertex-cover problem.
A simple greedy algorithm which picks an edge $(u,v)$ not yet covered by the candidate vertex cover
$V$ and  adds $u$ and $v$ to the vertex cover (of course removing all edges that have one end as $u$ 
or $v$ from the candidate edges) has an approximation ratio of $2$ 
(i.e. the vertex cover found using greedy algorithm is within a factor of
$2$ of the vertex cover of minimum size). It will be interesting to explore the approximation
ratio of our greedy algorithm. In general, finding algorithms for \textit{ApiPP} that have a \textit{good approximation
ratio} is an interesting direction. We suspect that we will have to utilize properties of the uncertainity
function $U$ for this purpose.


\paragraph{Connection to fairness in ML.}
"Proxy problems" discussed in this paper also appear in other contexts, such as \textit{fairness in machine learning (ML)}. 
In fairness in machine learning, the proxy problem corresponds to whether there is a set of features 
that can predict well the sensitive attributes (e.g. gender and race). Therefore, even if sensitive attributes
are removed from the training dataset, the ML algorithm can pick up on the proxies for these sensitive attributes
and lead to unfair algorithms~\cite{arxiv-1810.07155}. Exploring connections between proxy problems in different contexts is an interesting area of future research.

\paragraph{Orthogonal API design.} 
An related problem is whether it is possible to design APIs not to have proxies or to minimize the number of API proxies. This would add to the list of requirements for the API-design process, further complicating this optimization problem. While the NP-completeness of \textit{ApiPP} indicates that it is no feasible to take a design-then-check approach, a co-design approach to APIs that applies mechanisms such as inference privacy may be the way forward. An adjacent problem is that of extending an API surface, where the question is whether adding one new API to an existing set introduces new API proxies.



\bibliographystyle{abbrv}
\bibliography{references}

\end{document}